\documentclass[aps,pra,reprint,amsmath,amssymb,superscriptaddress,nobibnotes,showpacs]{revtex4-1}

\usepackage{graphicx}
\usepackage{amsmath}
\usepackage{gensymb}
\usepackage{upgreek}
\usepackage{tabularx}
\usepackage{tabularx,color}
\usepackage{multirow}
\usepackage{hyperref}

\makeatletter

\newcommand{\Rmnum}[1]{\expandafter\@slowromancap\romannumeral #1@}
\makeatother

\begin{document}

\title{Generation of a non-zero discord bipartite state with classical second-order interference}

\author{Yujun Choi}
\affiliation{Center for Quantum Information, Korea Institute of Science and Technology (KIST), Seoul, 02792, Republic of Korea}
\affiliation{Department of Physics, Yonsei University, Seoul, 03722, Republic of Korea}

\author{Kang-Hee Hong}
\affiliation{Department of Physics, Pohang University of Science and Technology (POSTECH), Pohang, 37673, Republic of Korea}

\author{Hyang-Tag Lim}
\affiliation{Department of Physics, Pohang University of Science and Technology (POSTECH), Pohang, 37673, Republic of Korea}
\affiliation{Present address : Institute of Quantum Electronics, ETH Zurich, Zurich CH-8093, Switzerland}

\author{Jiwon Yune}
\affiliation{Center for Quantum Information, Korea Institute of Science and Technology (KIST), Seoul, 02792, Republic of Korea}

\author{Osung Kwon}
\affiliation{Center for Quantum Information, Korea Institute of Science and Technology (KIST), Seoul, 02792, Republic of Korea}
\affiliation{Present address : National Security Research Institute, Daejeon, 34044, Republic of Korea}

\author{Sang-Wook Han}
\affiliation{Center for Quantum Information, Korea Institute of Science and Technology (KIST), Seoul, 02792, Republic of Korea}

\author{Kyunghwan Oh}
\affiliation{Department of Physics, Yonsei University, Seoul, 03722, Republic of Korea}

\author{Yoon-Ho Kim}
\affiliation{Department of Physics, Pohang University of Science and Technology (POSTECH), Pohang, 37673, Republic of Korea}

\author{Yong-Su Kim}\email{yong-su.kim@kist.re.kr}
\affiliation{Center for Quantum Information, Korea Institute of Science and Technology (KIST), Seoul, 02792, Republic of Korea}
\affiliation{Department of Nano-Materials Science and Engineering, Korea University of Science and Technology, Daejeon, 34113, Republic of Korea}

\author{Sung Moon}
\affiliation{Center for Quantum Information, Korea Institute of Science and Technology (KIST), Seoul, 02792, Republic of Korea}

\date{\today} 

\begin{abstract}
\noindent	
We report investigation on quantum discord in classical second-order interference. In particular, we theoretically show that a bipartite state with $D=0.311$ of discord can be generated via classical second-order interference. We also experimentally verify the theory by obtaining $D=0.197\pm0.060$ of non-zero discord state. Together with the fact that non-classicalities originated from physical constraints and information theoretic perspectives are not equivalent, this result provides an insight to understand the nature of quantum discord.\end{abstract}

\maketitle

\section{Introduction}

Quantum correlation plays essential roles in many quantum information processes as well as fundamental experiments. Quantum entanglement has been spotlighted as a representative example of quantum correlation~\cite{horodecki09}. However, it has been known that entanglement is not the only kind of quantum correlation~\cite{bennett99, horodecki05, NisetCerf06}. In order to capture all the nonclassical correlation, Ollivier and Zurek introduced a measure, {\it quantum discord}, from the information-theoretic perspective~\cite{OllivierZurek01}.

Quantum discord covers a broader concept of quantum correlation than entanglement, and thus it has entanglement as a subset. In other words, all the entangled states have non-zero discord, however, there exist non-zero discord states which are separable~\cite{modi12}. Quantum discord has been actively studied due to the extensive coverage beyond entanglement. It has been applied in various research fields such as quantum communication~\cite{dakic12,pirandola14}, quantum computation~\cite{datta08, lanyon08}, quantum metrology~\cite{modi11}, thermodynamics of information~\cite{zurek03, BrodutchTerno10}, and dynamics of open system~\cite{ShabaniLidar09, mazzola10}. There are also several studies on the generation of quantum discord in continuous variable regime~\cite{blandino12,hosseini14,chille15} and discrete variable regime~\cite{lanyon08,xu10,almeida14}.

In the sense that interference is a fundamental phenomenon which may induce some correlations, it would be of interest to investigate quantum discord with the context of interference. \textcolor{black}{Indeed, studying quantum discord in the context of interference may provide distinct quantumness criteria of physical constrains and information theoretic perspectives~\cite{ferraroparis12}.} There are a few studies that relate interference to discord~\cite{meda13,girolami14, zhang15}. In these works, the authors showed that the interference between Gaussian states can be revealed with the aid of an ancillary Gaussian discordant state~\cite{meda13}, and the discord can be a resource of estimating parameter in metrology~\cite{girolami14}. The relation between discord and interference for mixed states also was investigated~\cite{zhang15}. However, there is no study that directly shows the role of interference for generating discord.

In this paper, we study quantum correlation of bipartite systems in the context of interference. In particular, we consider both entanglement and discord with classical second-order interference. The theoretical and experimental results show that the second-order interference can have an important role for generating non-zero discord bipartite states. Since interference is a fundamental phenomenon both in classical and quantum physics, linking quantum correlation with interference will provide \textcolor{black}{an insight into} quantum correlation.

\section{Theory}

\subsection{Quantum discord : the definition}

We briefly review the concept of quantum discord. In classical information theory, given two random variables, $A$ and $B$, we can define mutual information in two alternative ways:
\begin{equation}
I(A:B) = H(A) + H(B) - H(A,B),
\end{equation}
\begin{equation}
J(A:B) = H(A) - H(A|B).
\end{equation}
where $H(A) = -\sum_{a}P_{A=a}\log P_{A=a}$ is Shannon entropy, $H(A,B) = -\sum_{a,b}P_{A=a,B=b}\log P_{A=a,B=b}$ is joint entropy of A and B, and $H(A|B)=\sum_{b}P_{B=b}H(A|B=b)$ is conditional entropy of A given B with probability distribution $P$ and outcome $a (b)$ for $A (B)$. The marginal probability distributions, $P_{A}$ and $P_{B}$, are obtained from joint probability distribution, $P_{A,B}$, such that  $P_{A}=\sum_{b}P_{A,B=b}$ and $P_{B}=\sum_{a}P_{A=a,B}$. By applying Bayes rule, $P_{A|B=b}=P_{A,B=b}/P_{B=b}$, we can easily show that $H(A|B) = H(A,B)-H(B)$, and thus find out that the two definitions of mutual information are equivalent.

Considering the quantum version of mutual information, we replace the classical probability distributions with density matrices of two parties, $\rho_{A}$, $\rho_{B}$, and $\rho_{AB}$, and Shannon entropy with von Neumann entropy, $S(\rho_{A})=-{\rm Tr}_{A}\rho_{A}\log \rho_{A}$, where $\rho_{A}={\rm Tr}_{B}\rho_{AB}$ and $\rho_{B}={\rm Tr}_{A}\rho_{AB}$ are reduced density matrices of joint density matrix, $\rho_{AB}$. For instance, $I(\rho_{AB}) = S(\rho_{A}) + S(\rho_{B}) - S(\rho_{AB})$. The two definitions of mutual information are no longer same in quantum version due to the conditional entropy in $J(A:B)$. The quantum conditional entropy, $S(\rho_{A|B})$, is not directly formulated because the state of A can be affected by the measurement on B. Therefore, $S(\rho_{A|B})$ has to be defined according to the measurement on B such that $S(\rho_{A|B}) = S(\rho_{A|\{\Pi^{B}_{j}\}}) = \sum_{j}P_{j}S(\rho_{A|\Pi^{B}_{j}})$ where $\rho_{A|\Pi^{B}_{j}} = \Pi^{B}_{j}\rho_{AB}\Pi^{B}_{j}/{\rm Tr}_{A,B}\Pi^{B}_{j}\rho_{AB}$, $P_{j} = {\rm Tr}_{A,B}\Pi^{B}_{j}\rho_{AB}$, and $\{\Pi^{B}_{j}\}$ is a set of measurement on B. Since $J(\rho_{AB})$ is measurement-dependent in quantum physics, we express it as $J(\rho_{AB})_{\{\Pi^{B}_{j}\}} = S(\rho_{A}) - S(\rho_{A|B}) = S(\rho_{A}) - S(\rho_{A|\{\Pi^{B}_{j}\}})$.

Quantum discord is then given by the difference between two expressions of the mutual information in quantum version:
\begin{equation}
D(\rho_{AB})_{\{\Pi^{B}_{j}\}} = I(\rho_{AB}) - J(\rho_{AB})_{\{\Pi^{B}_{j}\}}
\end{equation}
As $I(\rho_{AB})$ and $J(\rho_{AB})_{\{\Pi^{B}_{j}\}}$ are total and classical correlations~\cite{HendersonVedral01}, respectively, we finally obtain measurement-independent quantum discord by maximizing $J(\rho_{AB})_{\{\Pi^{B}_{j}\}}$ over all possible measurement set of ${\{\Pi^{B}_{j}\}}$:
\begin{equation}
D(\rho_{AB}) = I(\rho_{AB}) - \max_{\{\Pi^{B}_{j}\}} J(\rho_{AB})_{\{\Pi^{B}_{j}\}}.
\end{equation}
\textcolor{black}{Note that we present the above expression with projective measurements since a projector provides the optimal measurement for the conditional entropy of two qubit state}~\cite{hamieh04}.  Moreover, it is worth noting that quantum discord for bipartite pure state is equivalent to the measure for entanglement~\cite{datta08, HendersonVedral01}. 

\begin{figure*}[t]
\includegraphics[width=6in]{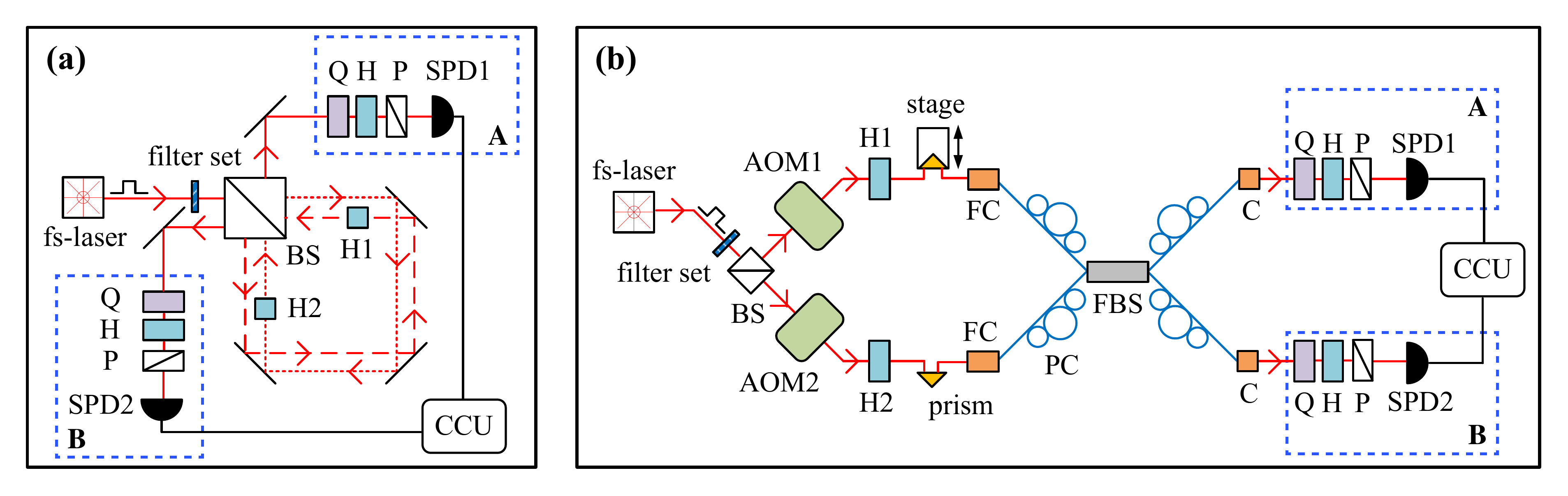}
\caption{Experimental setup where the phase of the two input pulses are mutually (a) coherent and (b) incoherent. fs-laser : femtosecond laser, filter set : filter set of neutral density filters and an interference filter, BS : 50:50 beamsplitter, AOM : acousto-optic modulator, FC : fiber coupler, PC : polarization controller, FBS : fiber beamsplitter, C : collimation lens, Q : quarter-wave plate, H : half-wave plate, P : polarizer, SPD : single photon detector, CCU : coincidence counting unit.}\label{setup}
\end{figure*}

\subsection{Quantum discord and \textcolor{black}{classical} second-order interference}

Let us consider a typical two-photon interference with a lossless symmetric beamsplitter (BS) that has two input modes $a, b$ and output modes $c, d$. The BS transformation between the input and output modes can be written with creation operators, $a^{\dag}\rightarrow\frac{1}{\sqrt{2}}(c^{\dag}+id^{\dag})$, and $b^{\dag}\rightarrow\frac{1}{\sqrt{2}}(d^{\dag}+ic^{\dag})$. With this BS transformation, we can think of various two-photon interferences by changing the input states.

First, let us investigate the case that two identical single photons are interfering at the BS, which corresponds to a typical Hong-Ou-Mandel interference~\cite{hom}. In this case, the coincidence between outputs, $c$ and $d$, becomes null, showing a Hong-Ou-Mandel (HOM) dip with an unit visibility, $V=1$. Note that the visibility of a HOM dip is defined as the relative depth of the dip with respect to the non-interfering cases.

If we change the polarization states of the incoming photons orthogonal to each other, the HOM dip disappears. However, in this setup, we can get a maximally entangled two-qubit state in polarization mode by post-selecting the case only when there is one photon at each output mode~\cite{shih88}. It is remarkable that the second-order interference is responsible for the generation of the entangled two-qubit state although there is no HOM dip with orthogonally polarized input photons. If the arrival time difference of the input photons is larger than the coherence time of the single-photon states, the output two-qubit state becomes completely mixed state~\cite{lim11}. Note that a HOM dip with $V>0.5$ cannot be explained with classical theory, and thus it is considered as quantum second-order interference~\cite{rarity05}. Therefore, this result shows that entanglement can be generated by quantum interference. Since all the entangled states have non-zero discord, we can conclude that quantum interference can generate quantum discord as well.

Now, let us turn our interest to classical second-order interference. For example, we can think of the case that two identical but phase randomized laser pulses are incoming to the input modes. In this case, one can get the HOM dip with limited visibility of V$\leq$0.5. Note that the first-order interference would be washed out due to the randomized phase between the two inputs. This result can be completely explained by the classical theory of superposition of electromagnetic waves, and thus it is considered as classical interference~\cite{rarity05,kim13,kim14}. We can get two-qubit states in polarization mode by setting the polarization states of the inputs orthogonal to each other and post-selecting the case when there is only one photon at each output, as we did for the single-photon inputs.

We can classify the classical two-photon interference according to the existence of mutual coherence between the laser pulses on the two input modes, $a$ and $b$. Assuming that the polarization states of the input pulses are horizontal and vertical respectively, the two-qubit states of mutually coherent inputs $\rho_{coh}$ and mutually incoherent inputs $\rho_{incoh}$ after the BS are given as
\begin{equation}
\rho_{coh}(\phi) = \frac{1}{4}
 \begin{pmatrix}
  1 & ie^{-i\phi} & -ie^{-i\phi} & e^{-2i\phi} \\
  -ie^{i\phi} & 1 & -1 & -ie^{-i\phi} \\
  ie^{i\phi} & -1 & 1 & ie^{-i\phi} \\
  e^{2i\phi} & ie^{i\phi} & -ie^{i\phi} & 1
 \end{pmatrix}
\label{rhocoh}
\end{equation}
\begin{equation}
\rho_{incoh} = \frac{1}{4}
 \begin{pmatrix}
  1 & 0 & 0 & 0 \\
  0 & 1 & -1 & 0 \\
  0 & -1 & 1 & 0 \\
  0 & 0 & 0 & 1
 \end{pmatrix}
\label{rhoincoh}
\end{equation}
\noindent
where $\phi$ is the phase difference between the input modes. The basis set of the density matrices is \{$|HH\rangle$, $|HV\rangle$, $|VH\rangle$, $|VV\rangle$\}, where $H$ and $V$ are horizontal and vertical polarization state, respectively. Note that $\langle\rho_{coh}(\phi)\rangle_{\phi}=\rho_{incoh}$, where $\langle X \rangle_{\phi}$ represents the average of X over many events of randomly varying $\phi$. The detailed derivation can be found in Appendix A.

From a given two-qubit density matrix, we can estimate the amount of entanglement and quantum discord~\cite{wootters98,yune15}. Note that the amount of entanglement can be quantified by concurrence. Regardless of the phase difference $\phi$, the concurrence and quantum discord of $\rho_{coh}(\phi)$ are $C(\rho_{coh}(\phi))=D(\rho_{coh}(\phi))=0$, showing that $\rho_{coh}(\phi)$ does not have any quantum correlation. On the other hand, the concurrence and discord of $\rho_{incoh}$ are $C(\rho_{incoh})=0$ and $D(\rho_{incoh})=-\left(\frac{3}{4}\right)\log_2\left(\frac{3}{4}\right)\approx0.311$, respectively. This result shows that classical second-order interference can generate a certain type of quantum correlation that can be captured by quantum discord. It is interesting to note that the existence of mutual coherence between the two input modes degrades the quantum discord.

Let us consider the physics behind these results. Taking into account the theorem that nonclassical inputs are indispensable for generating entanglement with a BS~\cite{kim02,xiangbin02}, it is obvious that the concurrence is zero in both cases. Quantum discord is not simply explained, but can be understood by investigating the reduced single-qubit states. Once we trace out one of the two qubits of Eq.(\ref{rhocoh}) and (\ref{rhoincoh}), we obtain the reduced single-qubit density matrices for each output of the BS, $c$ and $d$. The reduced single-qubit density matrices are given as
\begin{equation}
\rho_{coh|c}(\phi) = \frac{1}{2}
 \begin{pmatrix}
  1 & -ie^{-i\phi}  \\
  ie^{i\phi} & 1
   \end{pmatrix},
 \end{equation}
 \begin{equation}
\rho_{coh|d}(\phi) = \frac{1}{2}
 \begin{pmatrix}
  1 & ie^{-i\phi}  \\
 -ie^{i\phi} & 1
  \end{pmatrix},
 \end{equation}
\begin{equation}
\rho_{incoh|c} = 
\rho_{incoh|d} = \frac{1}{2}
 \begin{pmatrix}
  1 & 0  \\
  0 & 1
 \end{pmatrix},
 \end{equation}
\noindent 
where $\rho_{coh|c(d)}(\phi)$ and $\rho_{incoh|c(d)}$ denote the reduced single-qubit density matrices at output $c(d)$ for mutually coherent and incoherent inputs, respectively. Note that $\rho_{coh|c}(0) =|R\rangle \langle R|$ and $\rho_{coh|d}(0)=|L\rangle \langle L|$ where $|R\rangle \equiv \frac{1}{\sqrt{2}}(|H\rangle+i|V\rangle)$ and $|L\rangle \equiv \frac{1}{\sqrt{2}}(|H\rangle-i|V\rangle)$. By changing $\phi$, one can get single-qubit states that reside on the equator of Bloch sphere. 

From the investigation of reduced single-qubit states, the role of the mutual coherence between the input modes becomes clear. When the two inputs are mutually coherent, both output single-qubit states become pure and orthogonal to each other. As a result, the overall two-qubit state can be represented as a product of two single-qubit states, $\rho_{coh}=\rho_{coh|c}\otimes\rho_{coh|d}$, and thus no correlation exists in the bipartite system. When input pulses are incoherently mixed at a BS, on the other hand, both of the reduced single-qubit states become completely mixed. In this case, the overall two-qubit state cannot be represented as a simple product of single-qubit states, and shows non-zero quantum discord. Therefore, we can infer that the non-zero quantum discord of mutually incoherent inputs arises from the statistical mixing process.

\begin{figure*}[t]
\includegraphics[width=6in]{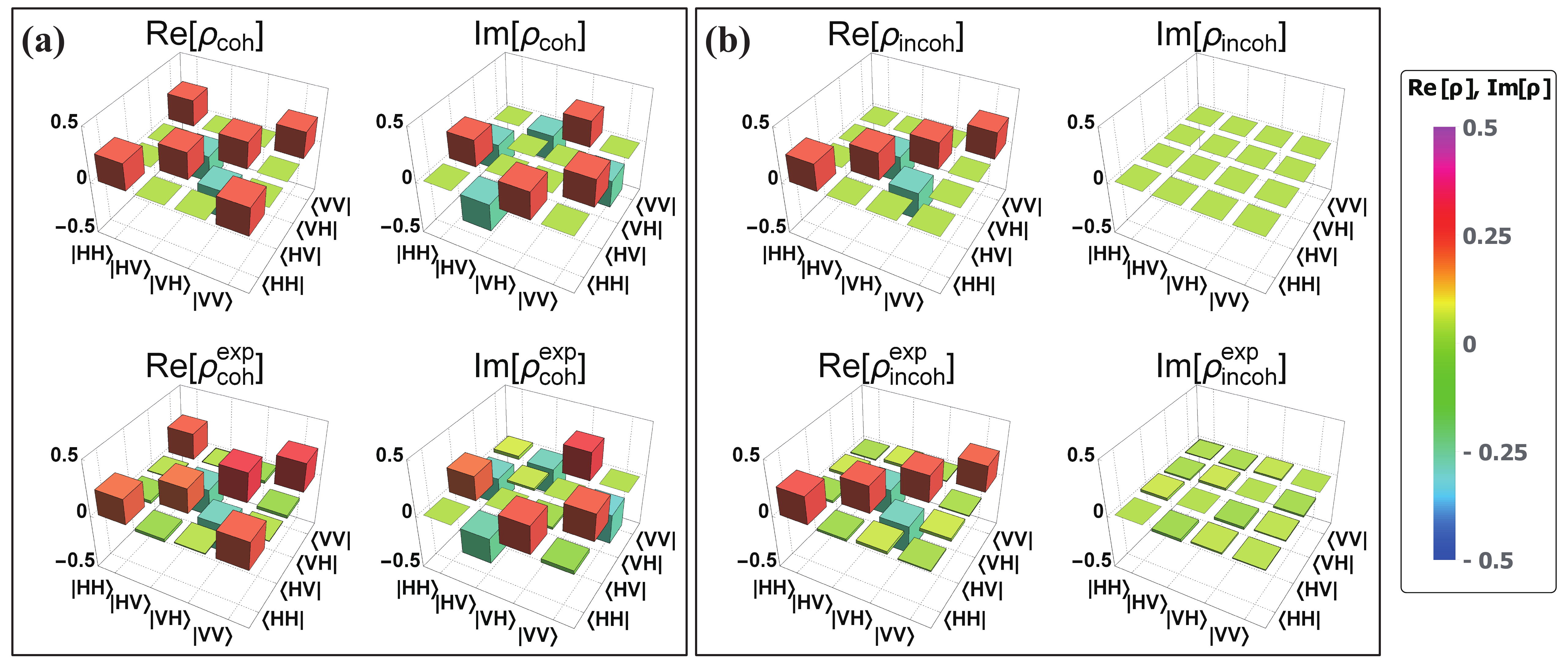}
\caption{Two-qubit density matrices where the phase of the pulses incident on the two input modes of a BS are mutually (a) coherent or (b) incoherent. $\rho_{coh}$ ($\rho_{coh}^{exp}$) is theoretical (experimental) density matrix of the coherent case. $\rho_{incoh}$ ($\rho_{incoh}^{exp}$) is theoretical (experimental) density matrix of the incoherent case. Re[$\cdot$] and Im[$\cdot$] denote real and imaginary part of the density matrix, respectively.}\label{DM}
\end{figure*}

\section{Experiment}

\subsection{Mutually coherent inputs}

In order to verify the theory, we construct an experimental setup as shown in Fig.~\ref{setup}(a) for the mutually coherent inputs case. A Ti:Sapphire mode-locked femtosecond laser (fs-laser) generates laser pulses whose temporal width is less than 200~fs and the repetition rate is about 80~MHz. The center wavelength of the pulses is tuned to 785~nm. The filter set consists of two parts: One is neutral density filters that attenuate the intensity of the pulses and the other is an interference filter whose  full width at half maximum (FWHM) of the transmission profile is about 3~nm.

To minimize the contribution of multi-photon states (more than two photons) to the coincidence detection, we attenuate the laser pulses so that the average photon number per pulse is about $\mu\approx0.1$. With this condition, the multi-photon states contribution is less than 10~\% of the total coincidence detection, see Appendix B for details.

After passing through the filters, the laser pulse enters a displaced Sagnac interferometer. Note that the second-order interference happens when two splitted laser pulses come out from the Sagnac interferometer. The Sagnac interferometer is employed to stabilize the mutual phase difference between the inputs. Although entanglement and discord are not affected by the phase difference, the stabilization of the phase difference during quantum state tomography is essential.  One can achieve high-level stabilization with the displaced Sagnac interferometer configuration~\cite{lee11,kim12}.

In the Sagnac interferometer, two half-wave plates (H1 and H2) are employed to make the polarization states of the input pulses orthogonal, $|H\rangle$ and $|V\rangle$. Note that H1 and H2 can be horizontally tilted to adjust the phase difference between the two inputs. During the experiment, the mutual phase difference is set at $\phi=0$. At the two outputs of the BS, two-qubit density matrix is reconstructed by means of quantum state tomography with maximum likelihood estimation using quarter-wave plates (Q), half-wave plates (H), and polarizers (P)~\cite{banaszek99,james01}. A home-made coincidence counting unit (CCU) based on an FPGA is used to register the coincidence counts between the SPDs~\cite{park15}. The coincidence counts are accumulated for 10 seconds for each projection state.

Figure~\ref{DM}(a) shows the theoretically (upper) and experimentally (lower) reconstructed two-qubit density matrices. The fidelity between these two states, $F=0.992\pm0.001$, shows that the experimentally reconstructed state is very close to the theoretical one. The concurrence and discord of the experimental two-qubit state is calculated as $C(\rho_{coh})=0.005\pm0.002$ and $D(\rho_{coh})=0.002\pm0.001$, which verifies that the state has neither entanglement nor quantum discord.

\subsection{Mutually incoherent inputs}

The experimental setup for the mutually incoherent inputs is depicted in Fig.~\ref{setup}(b). The light source and the tomographic setup (Q, H, P) are identical to those of the mutually coherent inputs case. Two asynchronous acousto-optic modulatrors (AOM1,2) are employed to wash out the first-order coherence between the two inputs~\cite{kim13, kim14}. To minimize the frequency shift by the AOMs, the operating RF frequencies of both AOM drivers are identically set to 40MHz. We block out all but the first-order diffracted pulses. H1 and H2 are employed to make the polarization states orthogonal to each other ($|H\rangle$ and $|V\rangle$). The pulses incident on fiber couplers (FC) interfere at a fiber beamsplitter (FBS). The polarization controllers (PC) at the input and output ports of the FBS compensate the polarization drift during fiber transmission. The output polarization output states are investigated with the following tomographic setup.

\begin{figure}[t]
\centering
\includegraphics[width=3.3in]{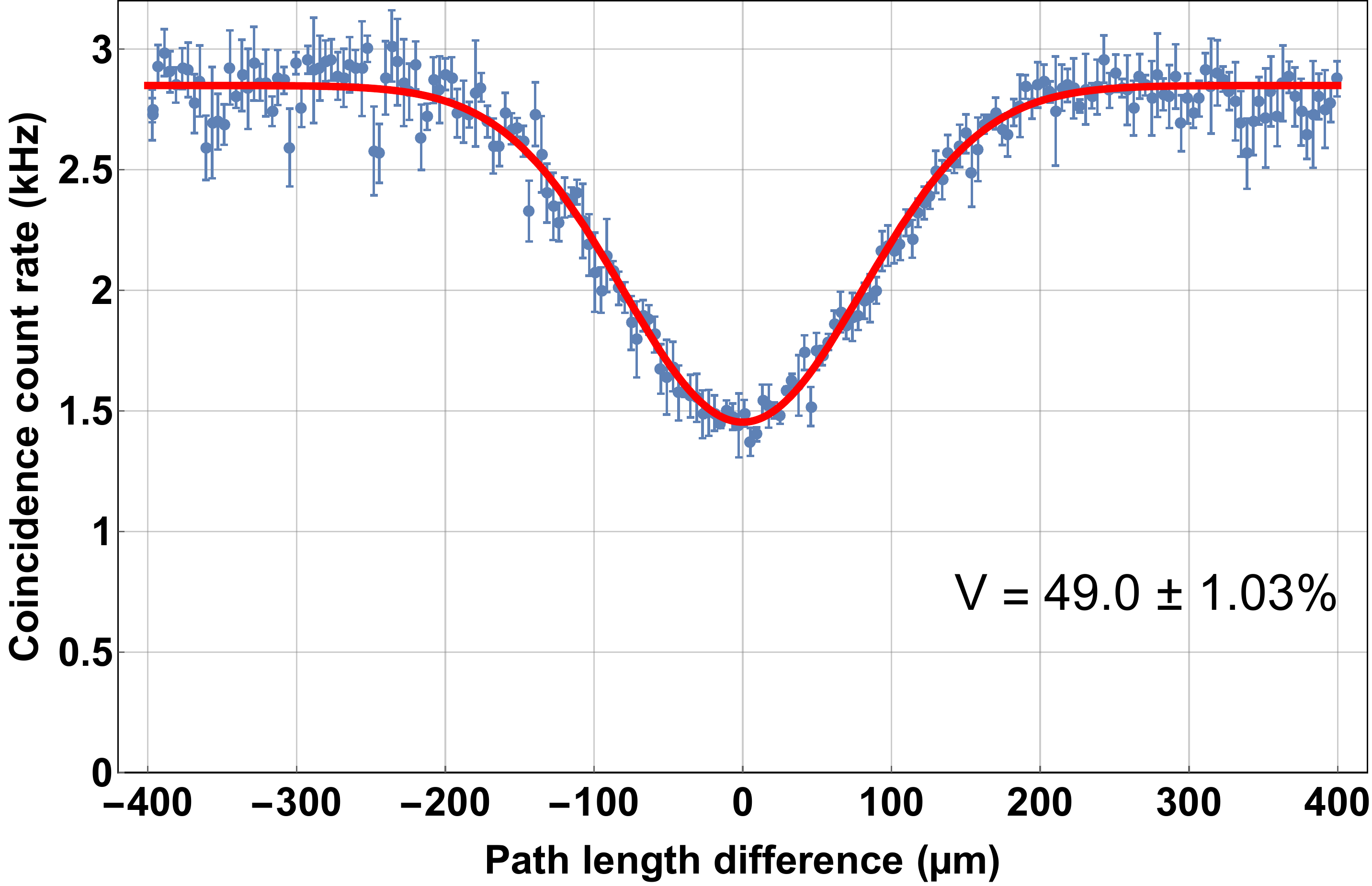}
\caption{Classical Hong-Ou-Mandel dip. The means and standard deviations of the coincidence count rate data are represented with blue circles and error bars, respectively. The red solid line denotes Gaussian fitting for the data. The visibility of the dip, $V$, is calculated from the fitted Gaussian function.}\label{HOMdip}
\end{figure}

Before we investigate the two-qubit state, we observe the classical second-order interference. To this end, we temporally set the polarization states of the inputs identical and remove tomographic setup. By scanning one of the optical paths with a translation stage, we measure the coincidences between the two outputs of the BS, see Fig.~\ref{HOMdip}. The $49.0\pm1.03\%$ visibility of classical HOM dip shows that spatial and polarization modes of the pulses are matched well. Note that the classical limit of HOM dip visibility is 50\%~\cite{rarity05}. The FWHM of the HOM dip, which is measured as 190~$\upmu{\rm m}$, also corresponds well to the coherence length elongated by the interference filter.

In order to get two-qubit state in polarization mode, we locate the stage at the zero position, make the input polarization states orthogonal, $|H\rangle$ and $|V\rangle$, and process the quantum state tomography. The theoretical and experimental density matrices are shown in Fig.~\ref{DM}(b). The concurrence and discord of this experimental two-qubit state are calculated as $C(\rho_{incoh})=0$ and $D(\rho_{incoh})=0.197\pm0.060$, which reveal that the two-qubit state is separable, however has non-zero discord. \textcolor{black}{It is remarkable that this result clearly shows that the criteria of quantumness originated from physical constraints and information theoretic perspectives do not coincide~\cite{ferraroparis12}}

\begin{figure}[t]
\centering
\includegraphics[width=3.3in]{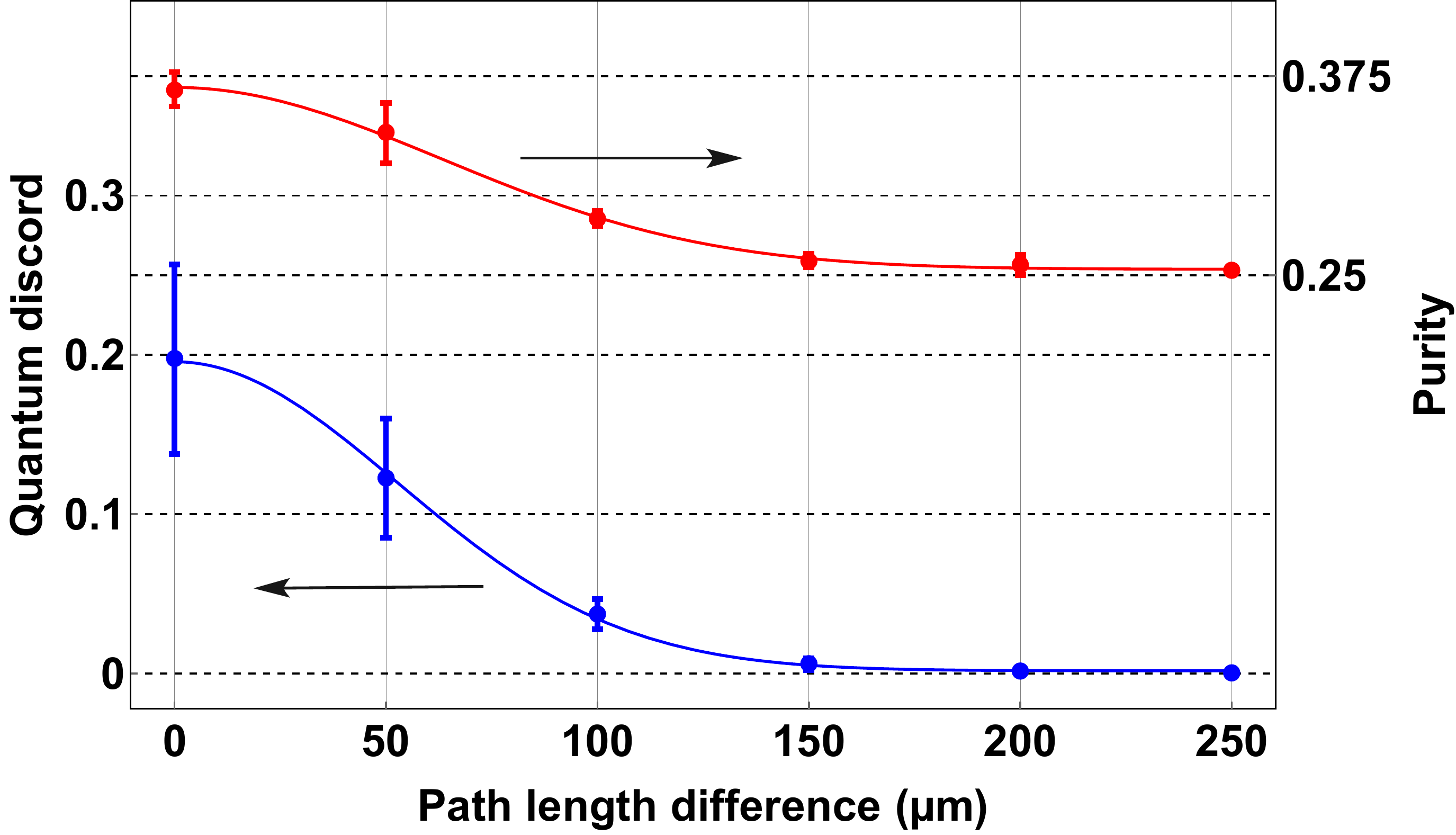}
\caption{Purity (red and right vertical axis) and discord (blue and left vertical axis) versus the optical path length difference for mutually incoherent case. The circles and error bars denote the experimental average values and standard deviations, respectively. The solid lines are Gaussian fittings for the data.}\label{DistanceIncoh}
\end{figure}

For the purpose of showing the role of the second-order interference, we experimentally investigate the purity and discord of the two-qubit states at various optical path length differences, see Fig.~\ref{DistanceIncoh}. As the optical path length difference between the two inputs increases, the purity and discord of the two-qubit state decrease. Note that the width of the decreasing discord with respect to the optical path length difference 125~$\upmu$m is comparable with the coherence length measured by the HOM dip in Fig.~\ref{HOMdip}. When the optical path length difference becomes larger than the coherence length, the two-qubit state asymptotically becomes completely mixed state which is a zero-discord state.

\section{Conclusion}

In conclusion, we theoretically and experimentally study entanglement and discord in classical second-order interference. When the laser pulses of the two input modes are mutually coherent, both entanglement and quantum discord of the post-selected output state become zero. On the contrary, when the pulses are mutually incoherent, the output state becomes separable, yet non-zero discord state. It shows that {\it classical} interference with post-selection can provide a certain type of {\it quantum} correlation, quantum discord. This result stimulates us to understand more about quantum correlation based on interference that is a fundamental phenomenon in both classical and quantum physics.

\section*{Acknowledgments}
\textcolor{black}{This work was supported by the ICT R$\&$D programs of MSIP/IITP (B0101-16-1355, R0101-15-0060), National Research Foundation of Korea (2013R1A2A1A01006029, 2015K2A1B8047228), and KIST Research Programs (2E26681, 2V05020).}

\newpage

\section{Appendix A : Two qubit states in classical second-order interference}

We develop the theory describing post-selected two-qubit state in classical second-order interference. We assume that the light incident on a lossless symmetric beamsplitter (BS) is monochromatic single-mode laser and the photon number of the output modes of the BS can be resolved. We post-select only the case where two photons are involved and one photon emerges from each ouput mode. The unitary transformation of the BS can be written as
 \begin{equation}
\begin{pmatrix}
a^{\dagger} \\
b^{\dagger}
\end{pmatrix} 
\to \frac{1}{\sqrt{2}}
 \begin{pmatrix}
1 & i \\
i & 1 \\
 \end{pmatrix}
\begin{pmatrix}
c^{\dagger} \\
d^{\dagger}
\end{pmatrix}
\label{bstrsf} 
\end{equation}
where $a^{\dagger}$ and $b^{\dagger}$ denote the creation operators at input modes, $c^{\dagger}$ and $d^{\dagger}$ are those for output modes. Note that the polarization states of the two input modes $a$ and $b$ are always orthogonal to each other. Therefore, we can simplify the situation as $a^{\dagger} \to a^{\dagger}_{H}$ and $b^{\dagger} \to b^{\dagger}_{V}$, where the subscripts $H$ and $V$ denote horizontal and vertical polarization, respectively.

\subsection{Mutually coherent inputs}

Since the photon number statistics of a laser follows Poissonian, we can write the laser pulse at each input mode of the BS as
\begin{equation}
|\alpha\rangle=e^{-|\alpha|^{2}/2}\sum\limits_{m=0}^{\infty} \frac{\alpha^{m}}{m!} (a^{\dagger}_{H})^{m} |0\rangle
\end{equation}
\begin{equation}
|\beta\rangle=e^{-|\beta|^{2}/2}\sum\limits_{n=0}^{\infty} \frac{\beta^{n}}{n!} (e^{i\phi}b^{\dagger}_{V})^{n} |0\rangle
\end{equation}
where $\phi$ is the phase difference between the two input modes. Assuming that the average photon numbers of the two input modes are the same, we can express $\alpha$ and $\beta$ as $\mu^{1/2}$ where $\mu$ is the average photon number. The overall input state of the BS can be represented as
\begin{equation}
|\psi\rangle_{coh}^{in}=e^{-\mu}\sum\limits_{m,n=0}^{\infty} \frac{\mu^{(m+n)/2}}{m!n!} (a^{\dagger}_{H})^{m} (e^{i\phi}b^{\dagger}_{V})^{n} |0\rangle|0\rangle.
\label{input}
\end{equation}

One can obtain the full expression of the output state with the BS transformation, Eq.~(\ref{bstrsf}), and the input state, Eq.~(\ref{input}) as follows. 
\begin{equation}
|\psi\rangle_{coh}^{out}=e^{-\mu}\sum\limits_{m,n=0}^{\infty} \frac{(\mu/2)^{(m+n)/2}}{m!n!} (c^{\dagger}_{H}+id^{\dagger}_{H})^{m} (e^{i\phi}(ic^{\dagger}_{V}+d^{\dagger}_{V}))^{n} |0\rangle|0\rangle.
\end{equation}
Since we are interested in such output state that each output mode has only one photon, the post-selected output state is given as
\begin{equation}
|\psi\rangle_{coh}=\frac{1}{2}(ic^{\dagger}_{H}d^{\dagger}_{H}+e^{i\phi}c^{\dagger}_{H}d^{\dagger}_{V}-e^{i\phi}c^{\dagger}_{V}d^{\dagger}_{H}+ie^{i2\phi}c^{\dagger}_{V}d^{\dagger}_{V}) |0\rangle|0\rangle.
\end{equation}
Note that this state is identical to the coherent output state, Eq.~(\ref{rhocoh}), i.e., $\rho_{coh}(\phi)=|\psi\rangle_{coh}\langle\psi|$.

\subsection{Mutually incoherent inputs}
When two input modes of the BS are mutually incoherent to each other, we can deal with the input state separately according to the incident photon numbers at each input mode, $i$ and $j$. Because we are interested in such output state that each output mode has only one photon, there are three cases for the input state contributing to the post-selected output state, $\rho_{incoh}$. The output state will be the statistical mixture of the output states of these three input states. In order to find the output density matrix, we calculate each case as follows.

\noindent
{\bf Case 1)} $|\psi\rangle_{(1,1)}^{in}=a^{\dagger}_{H}b^{\dagger}_{V}|0\rangle|0\rangle$. This input state corresponds to that each input mode has one photon, respectively. After the BS transformation, the output state is presented as
\begin{equation}
|\psi\rangle_{(1,1)}=\frac{1}{2}(ic^{\dagger}_{H}c^{\dagger}_{V}+c^{\dagger}_{H}d^{\dagger}_{V}-c^{\dagger}_{V}d^{\dagger}_{H}+id^{\dagger}_{H}d^{\dagger}_{V})|0\rangle|0\rangle.
\end{equation}
After the post-selection of one photon at each output mode, we can obtain the output state in the density matrix form as
 \begin{equation}
\rho^{(1,1)}_{incoh} = \frac{1}{2}
 \begin{pmatrix}
  0 & 0 & 0 & 0 \\
  0 & 1 & -1 & 0 \\
  0 & -1 & 1 & 0 \\
  0 & 0 & 0 & 0
 \end{pmatrix},
\end{equation}
which corresponds to a Bell state $|\psi^{-}\rangle$.

\noindent
{\bf Case 2)} $|\psi\rangle_{(2,0)}^{in}=\frac{1}{\sqrt{2}}(a^{\dagger}_{H})^2|0\rangle|0\rangle$. This input state corresponds to that there are two photons and no photon at input a and b, respectively. After the BS transformation, the output state is given as
\begin{equation}
|\psi\rangle_{(2,0)}=\frac{1}{2\sqrt{2}}((c^{\dagger}_{H})^2+2ic^{\dagger}_{H}d^{\dagger}_{H}-(d^{\dagger}_{H})^2)|0\rangle|0\rangle.
\end{equation}
Therefore, the post-selected output state can be presented in the density matrix form as
 \begin{equation}
\rho^{(2,0)}_{incoh} = 
 \begin{pmatrix}
  1 & 0 & 0 & 0 \\
  0 & 0 & 0 & 0 \\
  0 & 0 & 0 & 0 \\
  0 & 0 & 0 & 0
 \end{pmatrix}.
\end{equation}

\noindent
{\bf Case 3)} $|\psi\rangle_{(0,2)}^{in}=\frac{1}{\sqrt{2}}(b^{\dagger}_{V})^2|0\rangle|0\rangle$. This input state corresponds to that there are no photon and two photons at input a and b, respectively. After the BS transformation, the output state is given as
\begin{equation}
|\psi\rangle_{(0,2)}=\frac{1}{2\sqrt{2}}(-(c^{\dagger}_{V})^2+2ic^{\dagger}_{V}d^{\dagger}_{V}+(d^{\dagger}_{V})^2)|0\rangle|0\rangle.
\end{equation}
Therefore, the post-selected output state can be presented in the density matrix form as
 \begin{equation}
\rho^{(0,2)}_{incoh} = 
 \begin{pmatrix}
  0 & 0 & 0 & 0 \\
  0 & 0 & 0 & 0 \\
  0 & 0 & 0 & 0 \\
  0 & 0 & 0 & 1
 \end{pmatrix}.
\end{equation}

The overall two-qubit state $\rho_{incoh}$ of Eq.~(\ref{rhoincoh}) is given as the statistical mixture of $\rho^{(1,1)}_{incoh}$, $\rho^{(2,0)}_{incoh}$, and $\rho^{(0,2)}_{incoh}$. Taking account of the photon number distribution of laser pulse, we can get the overall two-qubit state as
\begin{equation}
\begin{split}
\rho_{incoh}=\frac{p(1,1)\rho^{(1,1)}_{incoh}+p(2,0)\rho^{(2,0)}_{incoh}+p(0,2)\rho^{(0,2)}_{incoh}}{{\rm Tr}[p(1,1)\rho^{(1,1)}_{incoh}+p(2,0)\rho^{(2,0)}_{incoh}+p(0,2)\rho^{(0,2)}_{incoh}]} \\
\end{split}
\end{equation}
where $p(1,1)=P(\mu,1)^2/2$, $p(2,0)=P(\mu,2)P(\mu,0)/2$, and $p(0,2)=P(\mu,0)P(\mu,2)/2$ are the conditional probabilities that correspond to $\rho^{(1,1)}_{incoh}$, $\rho^{(2,0)}_{incoh}$, and $\rho^{(0,2)}_{incoh}$, respectively. Since the $P(\mu,n)=e^{-\mu}\mu^n/n!$ is the Poissonian distribution of the photon numbers where $\mu$ and $n$ is the average photon number and the number of photons, the conditional probability $p(i,j)$ means that the probability of occurring coincidence event when $i$ and $j$ photons are incident on $a$ and $b$ input, respectively. 


\begin{figure}[t]
\centering
\includegraphics[width=3.3in]{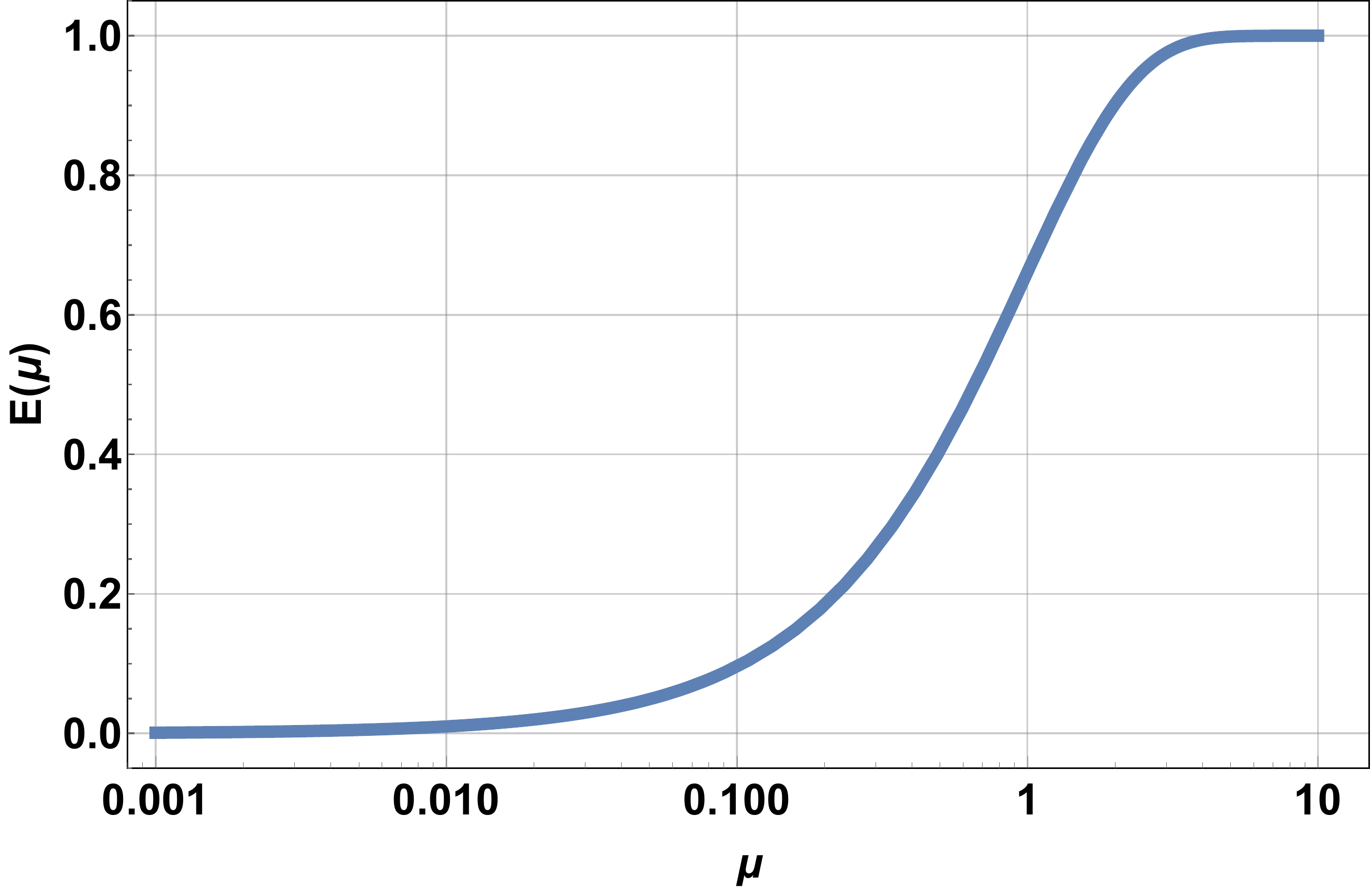}
\caption{The proportion of erroneous coincidence detections due to multi-photons.}\label{multi}
\end{figure}

\section{Appendix B : Multi-photon states contribution to coincidence detection}

Although we assume ideal single-photon detectors that resolve the number of photons in theory, practical single-photon detectors cannot resolve the photon numbers, and thus the multi-photon states can affect the coincidence detections. The multi-photon effect can be estimated by investigating the ratio of coincidence detections between two-photon states and multi-photon states. 

To this end, let us assume that the efficiencies of single photon detectors are not dependent on the number of photons incident on the detectors. Then, we can define an error function with respect to average photon number $\mu$, which estimates the proportion of erroneous coincidence detections to the total coincidence detections:
\begin{equation}
E(\mu) = \frac{\sum_{i+j \geq 3}{p(\mu,i,j)}}{\sum_{i+j \geq 2}{p(\mu,i,j)}}
\end{equation}
where $p(\mu,i, j)=P(\mu, i)P(\mu, j)Q(i+j)$ is conditional probability, $P(\mu,n) = e^{-\mu}\mu^{n}/n!$ is the Poisson distribution of the photon numbers, and $Q(n)=1-(\frac{1}{2})^{n-1}$ is the probability function of the number of photons involved in the coincidence detection. 

Figure~\ref{multi} shows $E(\mu)$ as a function of the average photon number $\mu$. Note that we limit the number of photons involved ($i+j$) to 100. As the average photon number increases, the proportion of erroneous coincidence detection due to multi-photon states increases. For $\mu=0.1$, which is investigated for our experiment, $E({\mu})=0.096$, and therefore we can conclude that most of the coincidence detections are due to the two-photon states.



\end{document}